\documentstyle[times,pramana,epsf,floats]{ias}
\begin{document}
\mark{{Probabilistic signatures of spatiotemporal intermittency in the coupled sine circle map lattice}{Zahera Jabeen and Neelima Gupte}}
\title{Probabilistic signatures of spatiotemporal intermittency in the coupled sine circle map lattice}
\author{Zahera Jabeen}
\address{Institute of Mathematical Sciences, Chennai, India.}
\author{Neelima Gupte}
\address{Department of Physics, Indian Institute of Technology-Madras, India.}
\keywords{Coupled map lattice, Spatiotemporal intermittency, Cellular automata}
\pacs{05.45.Ra, 05.45.-a, 05.45.Df, 64.60.Ak}

\abstract{

The phase diagram of the coupled sine circle map lattice 
exhibits a variety of interesting phenomena including spreading
regions with  spatiotemporal 
intermittency, non-spreading regions with spatial intermittency,
and coherent structures termed solitons. A cellular automaton mapping of
the coupled map lattice maps  the spreading to non-spreading
transition to a transition from a probabilistic to a deterministic
cellular automaton. The solitonic sector of the map shows spatiotemporal 
intermittency with soliton creation, propagation and annihilation. 
A probabilistic cellular automaton mapping is set up for this sector which can identify   
each one of these phenomena.
}

\maketitle

\section{Introduction}

The coupled sine circle map lattice has been known to model the
mode-locking behaviour \cite{gauri1} seen commonly in coupled
oscillators, Josephson Junction arrays, etc, and is also found to be
amenable to analytical studies \cite{Nandini,Nandini1}. 
In addition to mode-locked and synchronised behaviour, the model exhibits a rich variety
of phenomena such as spatio-temporal intermittency of the directed
percolation class \cite{Janaki,zjngpre1}
, spatial intermittency, cluster solutions and  
regimes with long lived coherent structures, also known as solitons
\cite{zjngpre2}. 

The coupled sine circle map lattice is defined on a one-dimensional
lattice, and evolves via the  equation:

\begin{equation}
x_i^{t+1}=(1-\epsilon)f(x_i^t)+\frac{\epsilon}{2}f(x_{i-1}^t) + \frac{\epsilon}{2}f(x_{i+1}^t) \pmod{1}\label{evol}
\end{equation}
Here, $x_i^t$ defines the state variable at site $i$ ($i=1,\ldots, N$, where $N$ is the system size) and time $t$. The site $i$ at each time step is coupled to its two nearest neighbours $i-1$ and $i+1$ with a coupling strength $\epsilon$. The local map at each site $i$, $f(x_i^t)$ is the sine circle map defined as $f(x)=x+\Omega-\frac{K}{2\pi}\sin(2\pi x)$, where $K$ denotes the strength of nonlinearity in the map, $\Omega$ is the frequency of the map in the absence of nonlinearity. 

\begin{figure}[!t]                                            
\epsfxsize=11cm                                               
\centerline{\epsfbox{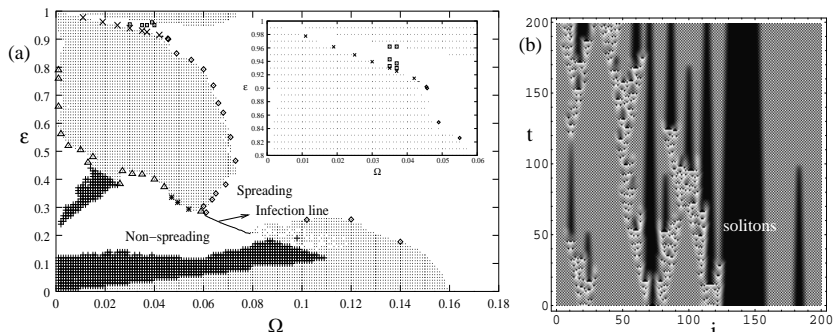}}                             
\caption{shows (a) the phase diagram obtained for a lattice of size
$N=1000$, after discarding $15, 000$ transients. Synchronised solutions
are seen in the region marked with dots. Spatiotemporal intermittency of
the DP class is seen at points marked with diamonds ($\Diamond$), and
spatial intermittency at points marked with triangles ($\triangle$) and
asterisks ($\ast$). Inset shows a part of the phase diagram where
spatiotemporal intermittency with travelling wave laminar states and
solitons is seen at points marked with boxes ($\Box$). (b) shows the
space-time plot of spatiotemporal intermittency with
solitons\label{pd}.}

\end{figure}

 The phase diagram of the model has been obtained in earlier studies
\cite{zjngpre1} by synchronously updating the coupled sine circle map lattice with random initial conditions, in the parameter region $0<\Omega\leq \frac{1}{2\pi}$, $0\leq\epsilon\leq1$ and at $K=1.0$. The phase diagram is seen to be organised around the bifurcation boundary of the spatiotemporally synchronised solution as well as a line, called the infection line, that separates the lower half of the parameter space into a spreading and a non-spreading regime (Figure \ref{pd}(a)). The burst states were seen to be capable of infecting their neighbouring laminar states in the spreading regime, whereas they were seen to be non-infectious and localised in the non-spreading regime.
Spatio-temporal intermittency contaminated by long lived travelling
coherent structures, also named as solitons,  was seen in the upper part
of the phase diagram (Figure \ref{pd}(a) inset).

 Special behaviour was seen near the bifurcation boundary of the synchronised solutions in these two regimes. In the spreading regime, spatiotemporal intermittency with exponents of the directed percolation (DP) universality class was seen near the bifurcation boundary of the synchronised solutions. This type of intermittency is completely free of coherent structures and a complete set of exponents obtained at the onset of this intermittency matches convincingly with the directed percolation class \cite{zjngpre1}. In the non-spreading regime, spatial intermittency (SI) with synchronised laminar states and periodic or quasi-periodic bursts  was seen near the bifurcation boundary \cite{zjngpre2}. The scaling exponents for the laminar length distributions in this type of intermittency were seen to be similar to those obtained for spatial intermittency seen in the inhomogeneous logistic map lattice \cite{ashutosh}. 
 

Further insights into the spreading to non-spreading transition were obtained when the coupled map lattice was mapped onto a cellular automaton \cite{zjngphysica}. In the spreading regime, including the points where DP behaviour is seen, the probabilities associated with the cellular automaton update rules were seen to lie in the $(0,1)$ interval. In other words, the update rules obtained in the spreading regime were seen to be probabilistic. In contrast, the probabilities in the entire non-spreading regime were seen to be either zero or one and were as follows : $p_0=0.0, p_1=0.0, p_2=0.0, p_3=1.0, p_4=0.0, p_5=1.0$. Thus, the update rules in non-spreading regime were found to be deterministic. Therefore, the cellular automaton mapping of the coupled map lattice showed that the spreading regime can be mapped onto a set of probabilistic cellular automata whereas the entire non-spreading regime can be mapped to a deterministic cellular automaton.

Apart from the spatiotemporal intermittency of the directed percolation
class and the spatial intermittency seen in the lower part of the phase
diagram, spatiotemporal intermittency with travelling wave laminar
states interspersed with turbulent bursts is  seen in the upper part of
the phase diagram (See inset of Figure \ref{pd}(a)).  This kind of spatiotemporal intermittency contains coherent structures or 'solitons', which spoil the analogy with  directed percolation  and are responsible for non-universal exponents in this region \cite{zjngpre2}.  In this paper, we show that a cellular automaton can be designed for this type of spatiotemporal intermittency, which successfully picks up signatures of the 'solitons'. We discuss this in the following section.


\section{Spatiotemporal intermittency with travelling wave laminar states and solitons}

Spatiotemporal intermittency with travelling wave laminar states is seen
in the upper part of the phase diagram at points marked with boxes ($\Box$)  (Figure \ref{pd}(a)). 
In this type of intermittency, the lattice relaxes to the absorbing travelling wave laminar state asymptotically. The burst states are turbulent and lie in the interval $(0,1)$, and can spread through the lattice. Apart from the burst states, coherent structures, which have been called 'solitons', are seen in the travelling wave laminar background. These structures  have been marked in the space-time plot of this type of spatiotemporal intermittency in Figure \ref{pd}(b). As can be seen from the space-time plot, these solitons are generated from the turbulent burst states. Moreover, both left-moving and right-moving solitons occur in pairs in the lattice and eventually annihilate each other on collision. When these solitons collide, they either die down to the travelling wave laminar state or give rise to turbulent bursts. Therefore, the burst states are created in the lattice either by infection of a laminar site by a neighbouring burst site or by the annihilation of solitons. Regions of synchronised fixed point solutions, $x^{\star}$ are seen to exist between the left and right moving solitons, which perish with the annihilation of the solitons. The coherent structures seen in this type of STI show a strong resemblance to the 'solitons' seen in the Chat\'e-Manneville coupled map lattice \cite{chate}, where these solitons were responsible for spoiling the directed percolation behaviour \cite{grassberger,bohr}.

The solitons alter the dynamical behaviour of the system in this regime
in several significant ways. They 
spoil the analogy with DP behaviour by acting as an additional source of
turbulent bursts. They also introduce an new timescale in the system, namely the soliton
lifetime, which depends on $\Omega$ and $\epsilon$ and gives rise to
non-universal exponents.
 This can be seen in Table \ref{soldis} in which the scaling exponents
$\zeta$ associated with the distribution of laminar lengths $P(l)\sim
l^{-\zeta}$, obtained at various values of coupling strengths $\epsilon$
for $\Omega=0.035$ have been shown. 

The exponent $\zeta$ is seen to vary
from $1$ to $1.5$ for different values of $\epsilon$. The soliton
lifetimes also decrease with increase in coupling strength and give rise
to two distinct regimes. In regimes of long average soliton lifetimes, the distribution of soliton lifetimes shows a power-law behaviour whereas the distribution shows a peak with a characteristic time-scale ($\sim20$) in regimes of short soliton lifetimes. 
These varying average soliton lifetimes influence the
  extent of spreading in the lattice and therefore  lead to varying values for the  laminar length distribution exponents \cite{zjngpre2}. 
Thus, the creation, propagation and annihilation of solitons leads to  
significant changes in the statistical and dynamical behaviour of the system. 
 A cellular automaton mapping for this type of spatiotemporal
intermittency can be designed which mimics the dynamics observed in this
region \cite{chate,wolfram,domanykinzel}, and contains significant signatures of
these solitonic processes. We discuss this in the next section.

\begin{table}[!t]
\begin{tabular}{cc|ccc}
\hline
\multicolumn{2}{c|}{$\Omega=0.035$}&\multicolumn{3}{c}{Distribution of soliton lifetimes}\\
\hline
~~~$\epsilon$~~~ & $\zeta$ & Nature & Scaling exponent &$T_{max}$\\
\hline
~~~~0.933~~~~ & ~~~~1.53 $\pm$ 0.01~~~~ & Scales with a power-law & 1.14 & 19010 \\
~~~~0.943~~~~ & ~~~~1.40 $\pm$ 0.01~~~~ & $"$ & 1.35& 2481\\
~~~~0.950~~~~ & ~~~~1.17 $\pm$ 0.01~~~~ & $"$ & 1.64& 794\\
~~~~0.962~~~~ & ~~~~1.02 $\pm$ 0.01~~~~ & Peaked distribution with a power-law tail & 2.84 & 305\\
\hline
\end{tabular}

\caption{ shows the laminar length distribution exponent, $\zeta$ obtained for different values of the coupling strength, $\epsilon$ at $\Omega=0.035$ in the STI with TW laminar state and turbulent bursts regime. The table also shows the scaling exponents associated with the distribution of soliton lifetimes, and the maximum soliton lifetime observed, $T_{max}$. \label{soldis}}
\vspace{-.2in}
\end{table}


\section{A cellular automaton for spatiotemporal intermittency containing solitons}

 The cellular automaton is defined on a $(1+1)D$ lattice. 
There are four states in the solitonic region, the travelling wave
laminar state, the burst state, the soliton state and the fixed point
state. 
Therefore, the  variable $v_i^t$ at site $i$ and time $t$ is assigned a value $v_i^t=0$ if it exists in the travelling wave laminar state, $v_i^t=1$ for a burst state, $v_i^t=2$ for a solitonic state, and $v_i^t=3$ if it exists in the fixed point state, $x^{\star}$.  Since, the probability, that the site $i$ at time $t+1$ exists in the state $v_i^{t+1}$, depends on the states of the sites $i-1, i, i+1$ at time $t$ as defined in the evolution equation \ref{evol}, the update rules of the cellular automaton are given by the conditional probability $P(v_i^{t+1}|v_{i-1}^t,v_i^t,v_{i+1}^t)$.
 
  For the above defined CA, $4^3$ initial states are possible. After
considering the $4$ possible final states $v_i^{t+1}=0,1,2,3$, we can
define $4\times4^3$ update rules of the cellular automaton in this
region. However, the number of relevant update rules required to
understand the dynamics is much smaller, and can be reduced  by (i)
using the symmetry between the sites $i-1$ and $i+1$ as defined in the
evolution equation, and considering the states $P(v_i^{t+1}|001)$ and
$P(v_i^{t+1}|100)$(say) as equivalent, (ii) eliminating the states which
do not exist as neighbours (eg: the states $v_i^t=0$ and $v_i^t=3$ do
not exist as nearest neighbours, as the travelling wave state and the
fixed point state are never nearest neighbours). Finally,  we reduce the
number further by focusing on
states which appear more frequently in the system.  
  
The values of these probabilities can be estimated from the evolution of the CML equation \ref{evol} starting from random initial conditions.  Let $N_k$ be the number of sites $i$ at time $t$, which along with its neighbours $i-1$ and $i+1$ exist in the $k^{th}$  state at time $t$. If $N^0_k$, $N_k^1$, $N^2_k$, and $N^3_k$ be the fraction of $N_k$ sites, which, at time $t+1$, exist in $0$ (travelling wave), $1$ (burst), $2$ (soliton) and $3$ (fixed point $x^{\star}$) states  respectively such that $N_k=N_k^0+N_k^1+N_k^2+N_k^3$, then the probability $p_k^m$ is estimated using $p_k^m=N_k^m/N_k$, where $m=0,1,2,$ or $3$.

As a result, we get the following $17$ prominent initial configurations
that constitute at least $0.1\%$ of the total dynamics: $000$,
$(001,100)$, $010$, $(011,110)$, $101$,  $111$, $(002,200)$,
$(022,220)$, $(112,211)$, $(122,221)$, $222$, $(331,133)$, $(311,113)$,
$(332,233)$, $(322,223)$, $232$, and $333$. 
Further identification  of the most 
relevant update rules is carried out by concentrating on those
configurations which show a higher probability of transition to the
final states $v_i^{t+1}=0,1,2,$ or $3$.

\begin{figure}[!t]                                            
\epsfxsize=12cm
\centerline{\epsfbox{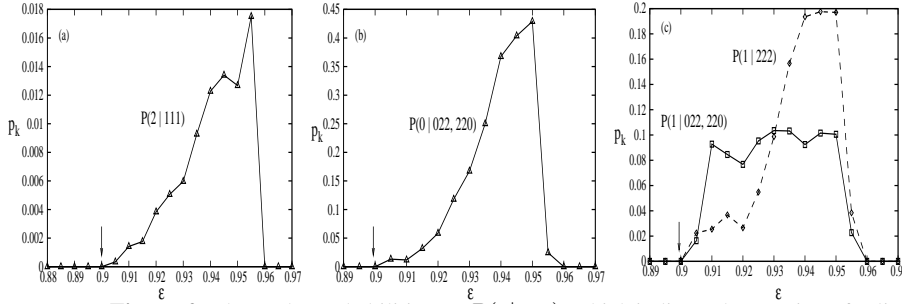}}
\caption{shows the probabilities (a) $P(2|111)$, which indicate the
creation of solitons, (b) $P(0|022,220)$ which is the probability of
propagation of solitons, and (c)$P(1|222)$ and $P(1|022, 220)$, which
gives the probability of creation of bursts due to the annihilation of
solitons. These probabilities have been obtained for various values of
the coupling strengths $\epsilon$ at $\Omega=0.0455$. }
\label{propann}                                               
\vspace{-.2in}
\end{figure}


The CA probabilities thus identified 
contain
information about the infective dynamics of the burst states and also 
show significant signatures of the solitons. For example, the
probabilities which are indicative of the infective behaviour of the
burst states include $P(1|001,100)$, $P(1|011,110)$, $P(1|101)$,
$P(1|111)$, $P(1|112,211)$, $P(1|122,221)$, and $P(1|133,331)$. In all
the above configurations, the variable at the central site $i$ changes
into a burst state at time $t+1$, after being infected by a neighbouring
burst state at time $t$.

Features of the creation, propagation and annihilation of solitons are
seen in the following update rules.
As discussed earlier, a small fraction of the burst states are seen to
generate pairs of left- and right-moving solitons. This information is
contained in the following probabilities: $P(2|001,100)$,
$P(2|011,110)$, $P(2|101)$, $P(2|111)$, $P(2|112,211)$, $P(2|133,331)$,
$P(2|113,311)$. For instance, the update rule defined by the conditional
probability $P(2|111)$ determines the probability of obtaining a
solitonic state ($v_i^{t+1}=2$) at site $i$ at time $t+1$, given that
the sites $i-1, i, i+1$ exist in the burst state
($v_{i-1}^t,v_i^t,v_{i+1}^t=1$) at time $t$. Figure \ref{propann}(a)
shows the probability $P(2|111)$ obtained at $\Omega=0.0455$ for various
values of the coupling strength $\epsilon$. The probability has been
calculated for a lattice of size $500$ over $5000$ time steps (after
discarding $3000$ transients), and has been averaged over $50$ initial
conditions. This soliton creation probability shows a prominent increase at
$\epsilon=0.9$, which implies that the burst states
  generate 'solitons' at $\epsilon>0.9$. This is also the point at which
spatiotemporal intermittency with synchronised laminar states bifurcates
to spatiotemporal intermittency with travelling wave laminar states and
solitons. The probabilities are further seen to die down to zero at
$\epsilon=0.96$. At this point, the synchronised solutions reappear in
the phase diagram.

The probabilities which are indicative of the propagation of solitons
through the lattice are as follows:  $P(0|022, 220)$, $P(2|022,220)$,
$P(2|122, 221)$, $P(2|222)$, $P(2|233,332)$, $P(2|223,322)$, $P(2|232)$.
Figure \ref{propann}(b) shows the probability $P(0|220,022)$ plotted for
various values of $\epsilon$ at $\Omega=0.0455$. The probability is
non-zero for $0.9<\epsilon<0.96$. And finally, when the left and
right-moving solitons collide with each other, they either die down to
the laminar state or give rise to turbulent bursts. Signatures of
annihilation of the solitons is seen in the probabilities:
$P(0|022,220)$, $P(1|022.220)$, $P(1|122,221)$, $P(1|222)$,
$P(1|223,322)$, and $P(1|232)$. The probabilities $P(1|220,022)$ and
$P(1|222)$ are plotted in Figure \ref{propann}(c). These probabilities
are again seen to be non-zero for $0.9<\epsilon<0.96$. Hence, we see
that the cellular automaton designed for spatiotemporal intermittency
with travelling wave states describes
  the infection dynamics of the burst states as well as picks up the
signatures of the creation, propagation and the annihilation of the
solitons.

\section{Conclusions}

In our earlier studies, we showed that the transition from the spreading
to non-spreading regime in the phase diagram of the coupled sine circle
map lattice can be mapped onto a cellular automaton, whereby the
cellular automaton undergoes a transition from a set of probabilistic
cellular automata in the spreading regime to a deterministic cellular
automaton in non-spreading regime. In this paper, we showed that a
similar cellular automaton can be designed for spatiotemporal
intermittency with travelling wave laminar states contaminated by
solitons. This cellular automaton effectively picks up the key
signatures of the solitons including their creation from the burst
states, propagation through the lattice and their annihilation. Moreover, it detects the regions in the phase diagram, where the solitons and spatiotemporal intermittency with travelling waves are seen to exist. 
Hence, we see that the cellular automaton mapping, though a
coarse-grained version of the coupled map lattice, can still recognise
the salient features of the local dynamics, while filtering out the
intricate details. Therefore, cellular automata mappings can be used as
a very useful tool in studying dynamics on spatially extended systems,
as well as in other areas such as  studies of critical behaviour of
absorbing phase transitions \cite{Odor,Odor1},  stock markets \cite{qiu}, traffic flows \cite{gao}, and cell growth kinetics \cite{block}.

\end{document}